\documentclass[preprint,showpacs,amsmath,amssymb,eqsecnum]{revtex4}
\usepackage{dcolumn}
\usepackage{bm}
\usepackage{graphicx}

\begin{document}
\draft
\title{{Large Lepton Mixing and 
Nonsymmetric Mass Matrices with Flavor 2 \(\leftrightarrow\) 3 Symmetry}}

\author{Koichi MATSUDA}
\affiliation{%
Graduate school of Science, 
Osaka University, Toyonaka, Osaka 560-0043, Japan}
\author{Hiroyuki NISHIURA}
\affiliation{%
Faculty of Information Science and Technology, 
Osaka Institute of Technology, 
Hirakata, Osaka 573-0196, Japan}

\date{June 19, 2005}

\begin{abstract}
We analyze the lepton sector of a recently proposed nonsymmetric mass matrix model. 
Our model gives a unified description of 
quark and lepton with the same texture form based on an extended flavor $2 \leftrightarrow 3$ symmetry with a phase.
By investigating possible types of assignment for masses, 
we find that the model can lead to large lepton mixings which are consistent with experimental values. 
We also find that the model predicts a very small value, $1.3\times10^{-10}$, 
for the lepton mixing matrix element square $|U_{13}|^2$. The $CP$ violating phases in the lepton mixing matrix 
and a suppression of the averaged neutrino mass in the neutrinoless double beta decay are also predicted.

\end{abstract}


\maketitle

\section{Introduction}
In order to explain a nearly bimaximal lepton mixing $(\sin^2 2\theta_{12}\sim 1$, 
$\sin^2 2\theta_{23}\simeq 1)$ observed from neutrino oscillation experiments~\cite{skamioka},   
mass matrices with various structures have been investigated in the literature. 
For example, mass matrices with texture zeros~\cite{fritzsch}-\cite{Ramond},  
with a flavor $2 \leftrightarrow 3$ symmetry~\cite{Fukuyama}-- \cite{Kaneko}, and so on have been proposed.  
Recently we have proposed~\cite{Matsuda3} a following nonsymmetric mass matrix model 
for all quarks and leptons based on an extended flavor $2 \leftrightarrow 3$ symmetry with one phase: 
\begin{equation}
M_f  =\left(
\begin{array}{lll}
\ 0 & \ a_fe^{-i\phi_f} & \ a_f \\
\ a^\prime_fe^{-i\phi_f} & \ b_fe^{-2i\phi_f} & \ (1-\xi_f)b_f \\
\ a^\prime_f & \ (1-\xi_f)b_f & \ b_f \\
\end{array}
\right) ,\quad \mbox{($f=u,d,\nu$, $e$, (\(D\), and \(R\)))} \label{ourMf}
\end{equation}
where $a_f$, $b_f$, $\xi_f$, and $a^\prime_f$ are real parameters and $\phi_f$ 
is a phase parameter. 
Here, \(M_u\), \(M_d\), \(M_\nu\), and  \(M_e\) are mass matrices
for up quarks (\(u,c,t\)), down quarks (\(d,s,b\)), 
neutrinos (\(\nu_e,\nu_\mu,\nu_\tau\)) and 
charged leptons (\(e,\mu,\tau\)), 
respectively.  
The mass matrices \(M_D\) and \(M_R\) are, respectively, 
the Dirac and the right-handed Majorana neutrino mass matrices, 
which are included in the model 
if we assume the seesaw mechanism \cite{Yanagida} for neutrino masses. 
In this model, we assume that all the mass matrices for quarks and leptons have this common structure, 
which is against the conventional picture that the mass matrix forms 
in the quark sector take somewhat different structures from those in the lepton sector.
In our previous works \cite{Matsuda3}, 
we have pointed out that this structure leads to reasonable values for 
the Cabibbo--Kobayashi--Maskawa (CKM) \cite{CKM} quark mixing, 
if we use a specific assignment of the quark masses. 
\par
In this paper, we shall discuss the Maki-Nakagawa-Sakata-Pontecorv (MNSP) \cite{MNSP} 
lepton mixing of the model by assuming that 
neutrinos are the Majorana particles. 
Under this assumption, the neutrino mass matrix \(M_\nu\) should be symmetric.
Namely, we further assume
\begin{equation}
a_{\nu}=a^\prime_{\nu}. 
\end{equation}
\par 
In the scenario that the neutrino mass matrix 
is constructed via the seesaw mechanism, 
i.e. \(M_\nu=-M_D^T M_R^{-1} M_D\),  
the structure of \(M_\nu\) mentioned above is 
alternatively realized by using the following two assumptions: 
(i) The mass matrices \(M_D\) and \(M_R\) have 
the same extended flavor 2 \(\leftrightarrow\) 3 symmetry in Eq.(\ref{ourMf})
with identical phase parameters, 
i.e. $\phi_D = \phi_R \equiv \phi_{\nu}$. 
(ii) \(M_D\) and \(M_R\) are proportional to each other, 
except for their (2,1) and (3,1) elements. 
\par
On the other hand,  \(M_e\) is assumed to have the above nonsymmetric structure given in Eq.~(\ref{ourMf}). 
Namely, in this paper, the mass matrices \(M_{e}\) and \(M_{\nu}\) are assumed to have the following forms:
\begin{eqnarray}
M_e  & =& \left(
\begin{array}{lll}
\ 0 & \ a_ee^{-i\phi_e} & \ a_e \\
\ a_e^\prime e^{-i\phi_e} & \ b_ee^{-2i\phi_e} & \ (1-\xi_e)b_e \\
\ a_e^\prime  & \ (1-\xi_e)b_e & \ b_e \\
\end{array}
\right) , \\
M_{\nu}  & =& \left(
\begin{array}{lll}
\ 0 & \ a_{\nu} e^{-i\phi_{\nu}} & \ a_{\nu} \\
\ a_{\nu} e^{-i\phi_{\nu}} & \ b_{\nu}e^{-2i\phi_{\nu}} & \ (1-\xi_{\nu})b_{\nu} \\
\ a_{\nu}  & \ (1-\xi_{\nu})b_{\nu} & \ b_{\nu} \\
\end{array}
\right) , 
\end{eqnarray}
where \(\phi_{e}\) and \(\phi_{\nu}\) are phase parameters. 

This article is organized as follows. 
In Sec.~II, we discuss the diagonalization of mass matrix of our model. 
The analytical expressions of the lepton mixings and phases of the model are given in Sec.~III. 
Sec.~IV  is devoted to a summary.

\section{Diagonalization of Mass matrix}
\par
\subsection{mass matrix for charged leptons }  
The diagonalization of the mass matrix for the charged leptons $M_e$ can be done similarly 
to the case of the mass matrices for up and down quarks. 
These mass matrices have common structure given by $M_f$ in Eq.~(\ref{ourMf}). 
Thus let us present a method for diagonalization of $M_f$, 
which is treated in Ref~\cite{Matsuda3} in details. 
\par
First, let us mention free parameters in the mass matrix.
There are five parameters, $a_f$, $a^\prime_f$, $b_f$, $\xi_f$, and $\phi_f$ in $M_f$. 
Even if we fix three eigenvalues, $m_{if}$, of $M_f$ by the observed fermion masses, 
there still exist two free parameters.  
As the free parameters, let us choose a parameter $\alpha_f$ defined by
\begin{eqnarray}
\alpha_f\equiv\frac{a_f^\prime}{a_f} ,\label{defalpha}
\end{eqnarray}
and a phase parameter $\eta_f$ defined in Fig.~1 independently of mass eigenvalues \(m_{if}\). 
Note that for the neutrino mass matrix $M_{\nu}$  we assume $\alpha_{\nu}=1$ as mentioned above. 
\par
Second, let us discuss an unitary matrix $U_{Lf}$ which diagonalizes $M_f M_f^\dagger$.
The explicit expression of $U_{Lf}$ depends on the following 
three types of assignment for $m_{if}$:
\par
(i) Type A:
\par
In this type, the mass eigenvalues \(|m_{1f}|\), \(m_{2f}\), and \(m_{3f}\) of $M_f$
are allocated to the masses of the first, second, and third generations, respectively.
(i.e. \(|m_{1f}|\)\(\ll\)\(m_{2f}\)\(\ll\)\(m_{3f}\).)
In this type, the $M_f M_f^\dagger$ is diagonalized  as
\begin{equation}
U_{Lf}^\dagger M_f M_f^\dagger U_{Lf} = \mbox{diag}\left(m_{1f}^2, m_{2f}^2, m_{3f}^2\right),
\end{equation}
by an unitary matrix $U_{Lf}$ given by 
\begin{equation}
U_{Lf}= P_{f}^\dagger O_f.\label{ul}
\end{equation}
Here $P_{f}$ is the diagonal phase matrix expressed as 
\begin{equation}
P_{f} =\mbox{diag}\left(1,e^{i(\phi_f-\varphi_f)},e^{-i\varphi_f}\right) \label{pf},
\end{equation}
where $\varphi_f$ and  $\phi_f$ are given by 
\begin{eqnarray}
\cos\varphi_f & =&\frac{|X_f|-m_{3f}\cos\eta_f}
			{\sqrt{|X_f|^2+m_{3f}^2-2m_{3f}|X_f|\cos\eta_f}} ,\label{eq3001}\\
\cos\phi_f & =&\frac{|X_f|^2-m_{3f}^2}
	{\sqrt{\left(|X_f|^2+m_{3f}^2\right)^2-4m_{3f}^2|X_f|^2\cos^2 \eta_f}} .\label{eq3003}
\end{eqnarray}
Here $X_f$ is defined by $X_f\equiv b_f+(1-\xi_f)b_fe^{i\phi_f}\equiv|X_f|e^{i\varphi_f}$, 
and $|X_f|$ is expressed in term of $\alpha_f$ and $m_{if}$ as
\begin{equation}
|X_f|^2  = m_{1f}^2+m_{2f}^2
			-|m_{1f}|m_{2f}\left(\frac{1+\alpha_f^2}{\alpha_f}\right). \label{Xf^2-A} 
\end{equation}
In Eq.~(\ref{ul}), $O_f$ is the orthogonal matrix given by
\begin{equation}
O_f\equiv
\left(
\begin{array}{ccc}
{  c_f}  & {  s_f}& {0} \\
{- \frac{s_f}{\sqrt{2}}} & {\frac{c_f}{\sqrt{2}}} & {-\frac{1}{\sqrt{2}}} \\
{- \frac{s_f}{\sqrt{2}}} & {\frac{c_f}{\sqrt{2}}} & {\frac{1}{\sqrt{2}}}
\end{array}
\right), \label{eq990114} 
\end{equation}
where 
\begin{equation}
c_f  =
\sqrt{\frac{m_{2f}^2 -\frac{|m_{1f}|m_{2f}}{\alpha_f}}{m_{2f}^2-m_{1f}^2}}\ ,\quad
s_f  =
\sqrt{\frac{\frac{|m_{1f}|m_{2f}}{\alpha_f}-m_{1f}^2}{m_{2f}^2-m_{1f}^2}}.\label{eq990115}
\end{equation}
It should be noted that the mixing angles are functions of only $\alpha_f$, 
since the $m_{if}$ is fixed by the experimental fermion mass values. 
We find from Eq.~(\ref{eq3003}) that $\phi_f \approx \pm \pi$ for 
$m_{1f}^2 \ll m_{2f}^2 \ll m_{3f}^2$ in this type A assignment. \\
\par
(ii)\ Type B: 
\par
In this type, the mass eigenvalues \(|m_{1f}|\), \(m_{3f}\), and \(m_{2f}\) 
are allocated to the masses of the first, second, and third generations, respectively.
(i.e. \(|m_{1f}|\)\(\ll\)\(m_{3f}\)\(\ll\)\(m_{2f}\).)
The $M_f M_f^\dagger$ is diagonalized as
\begin{equation}
U_{Lf}^\dagger M_f M_f^\dagger U_{Lf} = \mbox{diag}\left(m_{1f}^2, m_{3f}^2, m_{2f}^2\right).
\end{equation}
by an unitary matrix $U_{Lf}$ given by 
\begin{equation}
U_{Lf} = P_{f}^\dagger O_f^{\prime}.
\end{equation}
Here $O_f^{\prime}$ is obtained from \(O_f\) by exchanging the second row  
for the third one as  
\begin{equation}
O_f^\prime\equiv
\left(
\begin{array}{ccc}
{ c_f} & {0} & { s_f} \\
{- \frac{s_f}{\sqrt{2}}} & {\frac{1}{\sqrt{2}}} & {\frac{c_f}{\sqrt{2}}} \\
{- \frac{s_f}{\sqrt{2}}} & {-\frac{1}{\sqrt{2}}} & {\frac{c_f}{\sqrt{2}}}
\end{array}
\right). \label{eq990114B} 
\end{equation}
\par
(iii)\ Type C: 
\par
In this type, the mass eigenvalues \(m_{3f}\), \(|m_{1f}|\), and \(m_{2f}\) 
are allocated to the masses of the first, second, and third generations, respectively.
(i.e. \(m_{3f}\)\(\ll\)\(|m_{1f}|\)\(\ll\)\(m_{2f}\).) In this type, we have 
\begin{equation}
U_{Lf}^\dagger M_f M_f^\dagger U_{Lf} = \mbox{diag}\left(m_{3f}^2, m_{1f}^2, m_{2f}^2\right),
\end{equation}
where 
\begin{equation}
U_{Lf}=P_{f}^\dagger O_f^{\prime \prime}.
\end{equation}
Here, the orthogonal matrix \(O_f''\) is given by  
\begin{equation}
O_f''\equiv
\left(
\begin{array}{ccc}
{0}                   & { c_f}                   &  { s_f} \\
{\frac{1}{\sqrt{2}}}  & {- \frac{s_f}{\sqrt{2}}} &  {\frac{c_f}{\sqrt{2}}} \\
{-\frac{1}{\sqrt{2}}} & {- \frac{s_f}{\sqrt{2}}} &  {\frac{c_f}{\sqrt{2}}}
\end{array}
\right). \label{eq990114C} 
\end{equation}
This type is not so useful to get the reasonable lepton mixing values.

\subsection{mass matrix for neutrinos}
\par
Since the mass matrix for the Majorana neutrinos $M_{\nu}$ is symmetric,  $M_{\nu}$ is diagonalized as follows 
depending on the following three types of assignments for the neutrino mass $m_i$:\\
\par
(i) Type A:
\par
In this type, the mass eigenvalues \(m_{1}\), \(m_{2}\), and \(m_{3}\) of $M_{\nu}$
are allocated to the masses of the first, second, and third generations, respectively.
In this type, $M_\nu$ is diagonalized as
\begin{equation}
U_{\nu}^\dagger M_\nu U_{\nu}^* = \mbox{diag}\left(m_{1}, m_{2}, m_{3}\right),
\end{equation}
where $m_{i}(i=1,2,\mbox{and}\  3)$ are real positive neutrino masses. 
The unitary matrix $U_{\nu}$ is described as
\begin{equation}
U_{\nu} =P_{\nu}^\dagger O_\nu Q_{\nu}.
\end{equation}
Here, in order to make the neutrino masses $m_i$ to be real positive, 
we introduce a diagonal phase matrix $Q_{\nu}$ defined by 
\begin{equation}
Q_{\nu} \equiv \mbox{diag}\left(e^{-i(\varphi_\nu-\pi)/2},e^{-i(\varphi_\nu)/2},e^{-i(\varphi_\nu -\eta_\nu+\pi)/2}\right). \ \label{Q}
\end{equation}
The diagonal phase matrix $P_{\nu}$ and the orthogonal matrix $O_\nu$ are obtained 
from Eqs.~(\ref{pf}) -- (\ref{Xf^2-A}) and (\ref{eq990114}) -- (\ref{eq990115}) with $f=\nu$ by replacing \(|m_{1f}|\), \(m_{2f}\), and \(m_{3f}\) 
with \(m_{1}\), \(m_{2}\), and \(m_{3}\), respectively and by setting $\alpha_{\nu}=1$.\\

\par
(ii)\ Type B: 
\par
In this type, the mass eigenvalues \(m_{1}\), \(m_{3}\), and \(m_{2}\) 
are allocated to the masses of the first, second, and third generations, respectively.
In this type, $M_\nu$ is diagonalized as
\begin{equation}
U_{\nu}^\dagger M_\nu U_{\nu}^* = \mbox{diag}\left(m_{1}, m_{3}, m_{2}\right).
\end{equation}
The unitary matrix $U_{\nu}$ is described as
\begin{equation}
U_{\nu} =P_{\nu}^\dagger O_\nu^\prime Q_{\nu}^\prime.
\end{equation}
Here the diagonal phase matrix $Q_{\nu}^\prime$ is defined by 
\begin{equation}
Q_{\nu}^\prime \equiv \mbox{diag}\left(e^{-i(\varphi_\nu-\pi)/2},e^{-i(\varphi_\nu -\eta_\nu+\pi)/2},e^{-i(\varphi_\nu)/2}\right). \ \label{Q-B}
\end{equation}
The orthogonal matrix $O_\nu^\prime$ is obtained 
from Eqs.~(\ref{eq990114}) and (\ref{eq990115}) with $f=\nu$ by replacing \(|m_{1f}|\), \(m_{2f}\), and \(m_{3f}\) 
with \(m_{1}\), \(m_{3}\), and \(m_{2}\), respectively and by setting $\alpha_{\nu}=1$.\\

\par
(iii)\ Type C: 
\par
In this type, the mass eigenvalues \(m_{3}\), \(m_{1}\), and \(m_{2}\) 
are allocated to the masses of the first, second, and third generations, respectively.
In this type, $M_\nu$ is diagonalized as
\begin{equation}
U_{\nu}^\dagger M_\nu U_{\nu}^* = \mbox{diag}\left(m_{3}, m_{1}, m_{2}\right).
\end{equation}
The unitary matrix $U_{\nu}$ is described as
\begin{equation}
U_{\nu} =P_{\nu}^\dagger O_\nu^{\prime \prime} Q_{\nu}^{\prime \prime}.
\end{equation}
Here the diagonal phase matrix $Q_{\nu}^{\prime \prime}$ is defined by 
\begin{equation}
Q_{\nu}^{\prime \prime} \equiv \mbox{diag}\left(e^{-i(\varphi_\nu -\eta_\nu+\pi)/2},e^{-i(\varphi_\nu-\pi)/2},e^{-i(\varphi_\nu)/2}\right). 
\end{equation}
The orthogonal matrix $O_\nu^{\prime \prime}$ is obtained 
from Eqs.~(\ref{eq990114}) and (\ref{eq990115}) with $f=\nu$ by replacing \(|m_{1f}|\), \(m_{2f}\), and \(m_{3f}\) 
with \(m_{3}\), \(m_{1}\), and \(m_{2}\), respectively and by setting $\alpha_{\nu}=1$.\\

\par
These types B and C are not so useful to get the reasonable lepton mixing values.

\section{MNSP lepton mixing matrix}
\par
Now let us discuss the MNSP lepton mixing matrix 
of the model by taking the type A, the type B, and the type C assignments for charged 
leptons and neutrinos.  
We find that the assignment that is consistent with the present experimental data 
is only one case.
Namely, the case in which type B assignment for charged leptons and type A for neutrinos are taken. 
The other possible cases fail to reproduce consistent lepton mixing.
In this case, we obtain the MNSP lepton mixing matrix \(U\) as follows.
\begin{eqnarray}
U&=&U^\dagger_{Le}U_{\nu}=O^{\prime T}_eP_{e}P^\dagger_{\nu} O_{\nu}Q_{\nu}=O_e^{\prime T} P O_{\nu}Q_{\nu}\nonumber\\[.1in]
& =&
\left(
\begin{array}{ccc}
c^\prime_ec_{\nu}+\rho_{\nu} s^\prime_e s_{\nu} \quad & c^\prime_e s_{\nu}-{\rho}_{\nu} s^\prime_e c_{\nu} 
\quad & -{\sigma}_{\nu}s^\prime_e \\
{\sigma}_{\nu}s_{\nu} \quad & -{\sigma}_{\nu}c_{\nu} \quad & -\rho_{\nu} \\
s^\prime_e c_{\nu}-{\rho}_{\nu}c^\prime_e s_{\nu} \quad & s^\prime_e s_{\nu}+{\rho}_{\nu}c^\prime_e c_{\nu} 
\quad & {\sigma}_{\nu}c^\prime_e \\
\end{array}
\right)Q_{\nu},\label{ourmnsp-BA} 
\end{eqnarray}
where 
\begin{eqnarray}
s^\prime_e &= & \sqrt{\frac{\frac{|m_{e}|m_{\tau}}
	{\alpha_e}-m_{e}^2}{m_{\tau}^2-m_{e}^2}},\quad
c^\prime_e = \sqrt{\frac{m_{\tau}^2-\frac{|m_{e}|m_{\tau}}{\alpha_e}}
	{m_{\tau}^2-m_{e}^2}},  \nonumber \\
s_{\nu} &= &\sqrt{\frac{|m_{1}|}{m_{2}+|m_{1}|}},\quad
c_{\nu} = \sqrt{\frac{m_{2}}{m_{2}+|m_{1}|}}. 
\end{eqnarray}
Here the phase matrix $Q_{\nu}$ is shown in Eq.~(\ref{Q}), and we have put 
\begin{equation}
P \equiv P_{e}P^\dagger_{\nu} \equiv \mbox{diag}(1, e^{i\delta_{\nu 2}},e^{i\delta_{\nu 3}}) . 
\label{P}
\end{equation}
The orthogonal matrices $O_{\nu}$ and $O^\prime_e$ are obtained from Eq.~(\ref{eq990114}) and 
Eq.~(\ref{eq990114B}), respectively. 
Here we denote the lepton masses $(m_{1f},m_{3f},m_{2f})$ as $(m_e,m_{\tau},m_{\mu})$ for $f=e$, and 
as $(m_1,m_2,m_3)$ for $f={\nu}$. Note also that $\alpha_{\nu}=1$.  
\par
The parameters \(\rho_{\nu}\) and \(\sigma_{\nu}\) in Eq.~(\ref{ourmnsp-BA}) are defined by 
\begin{eqnarray}
\rho_{\nu} & =&\frac{1}{2}(e^{i\delta_{\nu 3}}+e^{i\delta_{\nu 2}})
=\cos\left(\frac{\delta_{\nu 3} - \delta_{\nu 2}}{2}\right) \exp i
\left( \frac{\delta_{\nu 3} + \delta_{\nu 2}}{2} \right) \ ,\label{rho}\\ 
\sigma_{\nu} & =&\frac{1}{2}(e^{i\delta_{\nu 3}}-e^{i\delta_{\nu 2}})
= \sin\left(\frac{\delta_{\nu 3} - \delta_{\nu 2}}{2}\right) 
\exp i \left( \frac{\delta_{\nu 3} + \delta_{\nu 2}}{2}+ \frac{\pi}{2}
\right) \ . \label{sigma}
\end{eqnarray}
Note that the phases of \(\rho_{\nu}\) and \(\sigma_{\nu}\) are 
\begin{eqnarray}
\mbox{arg} \ \rho_{\nu} 
	& =& \left\{
\begin{array}{ll}
	\frac{\delta_{\nu 3} + \delta_{\nu 2}}{2}
	&
	\mbox{\qquad for } \cos\left(\frac{\delta_{\nu 3} - \delta_{\nu 2}}{2}\right)>0 \\
	\frac{\delta_{\nu 3} + \delta_{\nu 2}}{2} +\pi
	& 
	\mbox{\qquad for } \cos\left(\frac{\delta_{\nu 3} - \delta_{\nu 2}}{2}\right)<0 \\
\end{array}\right. \ , 
\\ 
\mbox{arg} \ \sigma_{\nu}
	&=& \left\{
\begin{array}{ll}
	\frac{\delta_{\nu 3} + \delta_{\nu 2}}{2}+ \frac{\pi}{2}
	&
	\mbox{\qquad for } \sin\left(\frac{\delta_{\nu 3} - \delta_{\nu 2}}{2}\right)>0\\
	\frac{\delta_{\nu 3} + \delta_{\nu 2}}{2}- \frac{\pi}{2}.
	&
	\mbox{\qquad for } \sin\left(\frac{\delta_{\nu 3} - \delta_{\nu 2}}{2}\right)<0
\end{array}
\right. . 
\end{eqnarray}
By using Eqs.~(\ref{P}) and (\ref{pf}), the phases $\delta_{\nu 2}$ and $\delta_{\nu 3}$ 
in our model are given by
\begin{eqnarray}
\delta_{\nu 2} & =& \varphi_{\nu}-\varphi_e - (\phi_{\nu}-\phi_e), \label{delta2}\\
\delta_{\nu 3} & =& \varphi_{\nu}-\varphi_e . \label{delta3}
\end{eqnarray}
Here the phases $\phi_e$, $\varphi_e$, $\phi_{\nu}$, and $\varphi_{\nu}$ are expressed as 
\begin{eqnarray}
\cos \phi_e & =& \frac{|X_e|^2-m_{\mu}^2}
	{\sqrt{\left(|X_e|^2+m_{\mu}^2\right)^2-4m_{\mu}^2|X_e|^2\cos^2 \eta_e}} ,\label{phi-e-BA}\\
\cos \phi_{\nu} & =&  \frac{|X_{\nu}|^2-m_{3}^2}
	{\sqrt{\left(|X_{\nu}|^2+m_{3}^2\right)^2-4m_{3}^2|X_{\nu}|^2\cos^2 \eta_{\nu}}}  ,\label{phi-nu-BA}\\
\cos \varphi_e & =& \frac{|X_e|-m_{\mu}\cos \eta_e}
	{\sqrt{|X_e|^2+m_{\mu}^2-2m_{\mu}|X_e|\cos \eta_e}} ,\\
\cos \varphi_{\nu} & =&  \frac{|X_{\nu}|-m_{3}\cos \eta_{\nu}}
	{\sqrt{|X_{\nu}|^2+m_{3}^2-2m_{3}|X_{\nu}|\cos \eta_{\nu}}} ,
\end{eqnarray}
where 
\begin{eqnarray}
|X_e|^2 & =&m_{e}^2+m_{\tau}^2-|m_{e}|m_{\tau}\left(\frac{1+\alpha_e^2}{\alpha_e}\right) ,\\
|X_{\nu}|^2 & =&m_{1}^2+m_{2}^2-2|m_{1}|m_{2}
\label{X-nu-BA}. 
\end{eqnarray}
\par
In the following discussions we consider the normal mass hierarchy 
$|m_1| < m_2 \ll m_3$ for the neutrino masses.
Then the evolution effects can be ignored. 
Scenarios in which the neutrino masses have the quasi degenerate 
or the inverse hierarchy  
are denied from Eqs.~(\ref{eq30300}) and (\ref{eq20501}). 
\par
In order to reproduce the large lepton mixing between the second and third generation, 
we now choose specific values of the parameters $\alpha_e$ and $\eta_e$ such that
\begin{align}
&{\alpha_e}  
	= \frac{ m_e^2 - m_\mu^2 + m_\tau^2 + 
         {\sqrt{ {( m_e^2 - m_\mu^2 + m_\tau^2)}^2 - 4 m_e^2 m_\tau^2}} }
         {2 m_e m_\tau}
	\simeq \frac{m_{\tau}}{|m_e|}
	  \left[1-\left(\frac{m_{\mu}}{m_{\tau}}\right)^2\right], 
	  \label{alpha-e}\\
&\cos^2 \eta_e   \ne 1.
\end{align}
Then, we obtain 
\begin{align}
\varphi_e & \simeq \frac{\pi-\eta_e}{2} \ , 
&
\varphi_\nu & \simeq \pi-\eta_\nu \ , 
&
\phi_e & \simeq \left\{
	\begin{array}{ll}
		\frac{\pi}{2}  & \mbox{for  }  0 <\eta_e <\pi \\
		\frac{3\pi}{2} & \mbox{for  }  \pi <\eta_e <2\pi
	\end{array}\right.\ , 
& 
\phi_{\nu} &\simeq \pi,
\end{align}
and
\begin{align}
|\rho_\nu| & \simeq  \frac{1}{\sqrt{2}},
&
|\sigma_\nu| & \simeq \frac{1}{\sqrt{2}},
&
s_e^\prime & \simeq \frac{|m_e|m_\mu}{m_\tau^2}=1.63 \times 10^{-5},
&
c_e^\prime & \simeq 1.\
\end{align}
Thus, the explicit magnitudes of the components of $|U_{ij}|$ are obtained as
\begin{align}
\left|U_{11}\right|  & \simeq \sqrt{\frac{m_2}{m_2+m_1}}, & 
\left|U_{12}\right|  & \simeq \sqrt{\frac{m_1}{m_2+m_1}}, &
\left|U_{13}\right|  & \simeq \frac{1}{\sqrt{2}} \frac{|m_e|m_\mu}{m_\tau^2}, \nonumber \\
\left|U_{21}\right|  & \simeq \frac{1}{\sqrt{2}}\sqrt{\frac{m_1}{m_2+m_1}} , &
\left|U_{22}\right|  & \simeq \frac{1}{\sqrt{2}}\sqrt{\frac{m_2}{m_2+m_1}} , & 
\left|U_{23}\right|  & \simeq \frac{1}{\sqrt{2}},\nonumber \\
\left|U_{31}\right|  & \simeq \frac{1}{\sqrt{2}}\sqrt{\frac{m_1}{m_2+m_1}} , & 
\left|U_{32}\right|  & \simeq \frac{1}{\sqrt{2}}\sqrt{\frac{m_2}{m_2+m_1}} , &
\left|U_{33}\right|  & \simeq \frac{1}{\sqrt{2}}. \label{abs-u}
\end{align}
\par
In Fig.~2, we present more detailed predicted values for $|U_{23}|$ in the $\eta_e-\eta_{\nu}$ parameter space, 
by taking the value for $\alpha_e$ given in Eq.~(\ref{alpha-e}). 
It can be seen from Fig.~2 that the large mixing angle between the second and third generation is well realized in the model
if we use the specific values of $\alpha_e$ given in Eq.~(\ref{alpha-e}). 
\par
As seen from  Eq.~(\ref{abs-u}), the neutrino oscillation angles of the model are related to the lepton masses as follows: 
\begin{eqnarray}
\tan^2\theta_{\mbox{{\tiny solar}}}& =&\frac{|U_{12}|^2}{|U_{11}|^2}\simeq \frac{m_1}{m_2}\ ,\label{eq30300}\\
\sin^2 2\theta_{\mbox{{\tiny atm}}}& =&4|U_{23}|^2|U_{33}|^2\simeq 1\ ,
\label{eq30310} \\
|U_{13}|^2 &\simeq& \frac{1}{2} \left(\frac{m_e m_\mu}{m_\tau^2}\right)^2. \label{eq30320}
\end{eqnarray}
It should be noted that the present model leads to the same results for $\theta_{\mbox{{\tiny solar}}}$ and $\theta_{\mbox{{\tiny atm}}}$ as the model in Ref\cite{Matsuda2}, 
while a different feature for $|U_{13}|^2$ is derived.
\par
On the other hand, we have\cite{Garcia} a experimental bound for $|U_{13}|_{\mbox{\tiny exp}}^2$
from the CHOOZ\cite{chooz}, solar\cite{sno}, and atmospheric neutrino 
experiments\cite{skamioka}. 
From the global analysis of the SNO solar neutrino experiment\cite{sno,Garcia}, 
we have $\Delta m_{12}^2$ and $\tan^2 \theta_{12}$ for the large mixing angle (LMA) Mikheyev-Smirnov-Wolfenstein (MSW) solution.
From the atmospheric neutrino experiment\cite{skamioka,Garcia} , 
we also have $\Delta m_{23}^2$ and $\tan^2 \theta_{23}$. 
These experimental data with $3\sigma$ range are given by 
\begin{eqnarray}
& &|U_{13}|_{\mbox{\tiny exp}}^2 <  0.054 \ \label{mat20820} \ ,\\
& &\Delta m_{12}^2=m_2^2-m_1^2= \Delta m_{\mbox{{\tiny sol}}}^2
=(5.2-9.8) \times 10^{-5}\, \mbox{eV}^2, \label{mat20830} \\
& &\tan^2 \theta_{12}=\tan^2 \theta_{\mbox{{\tiny sol}}}=0.29-0.64 \ ,\label{eq20501}\\
& &\Delta m_{23}^2=m_3^2-m_2^2 \simeq \Delta m_{\mbox{{\tiny atm}}}^2
= (1.4-3.4) \times 10^{-3}\, \mbox{eV}^2, \label{mat20831}\\ 
& &\tan^2 \theta_{23} \simeq \tan^2 \theta_{\mbox{{\tiny atm}}}=0.49-2.2 \ . \label{mat208302}
\end{eqnarray}
Hereafter, for simplicity, we take  $\tan^2 \theta_{\mbox{{\tiny atm}}} \simeq 1$.
Thus, by combining the present model with the mixing angle \(\theta_{\mbox{{\tiny sol}}}\),
we have 
\begin{equation}
\frac{m_1}{m_2} \simeq \tan^2\theta_{\mbox{{\tiny sol}}}=0.29 - 0.64. 
\label{ratio}
\end{equation}
Therefore we predict the neutrino masses as follows.
\begin{eqnarray}
m_1^2 & = & (0.48-6.8) \times 10^{-5} \  {\rm eV^2} \ ,\nonumber \\
m_2^2 & = & (5.7-16.6) \times 10^{-5} \  {\rm eV^2} \ ,  \label{neu-mass}\\
m_3^2 & = & (1.4-3.4) \times 10^{-3} \  {\rm eV^2} \ .\nonumber
\end{eqnarray}
Let us mention a specific feature of the model. Our model imposes a stringent restriction on \(|U_{13}|\) as
\begin{equation}
|U_{13}|^2 \simeq \frac{1}{2} \left(\frac{m_e m_\mu}{m_\tau^2}\right)^2=1.3\times10^{-10} \ . \label{mat20870} 
\end{equation}
Here we have used the running charged lepton masses at the unification scale \(\mu=\Lambda_X\)
\cite{Fusaoka}: $m_e(\Lambda_X)=0.325\ \mbox{MeV}$, 
$m_\mu(\Lambda_X)=68.6\ \mbox{MeV}$, 
and $m_\tau(\Lambda_X)=1171.4 \pm 0.2\ \mbox{MeV}$.
The value in Eq.(\ref{mat20870}) is consistent with the present experimental constraints 
Eq.(\ref{mat20820}), however it is too small to be checked in neutrino factories in future.
The very small predicted value for $|U_{13}|$ is in contrast to previously proposed model\cite{Koide}\cite{Matsuda2}. 
\par
Next let us discuss the CP-violation phases in the lepton mixing matrix.
The Majorana neutrino fields do not have the freedom of rephasing 
invariance, so that we can use only the rephasing freedom of $M_e$ 
to transform Eq.~(\ref{ourmnsp-BA}) to the standard form
\begin{eqnarray}
& &U_{\rm std} 
= \mbox{diag}(e^{i\alpha_1^e},e^{i\alpha_2^e},e^{i\alpha_2^e}) 
\ U  \nonumber \\ 
& &= \left(
\begin{array}{ccc}
c_{\nu13}c_{\nu12} & c_{\nu13}s_{\nu12}e^{i\beta} & 
s_{\nu13}e^{i(\gamma-\delta_{\nu})} \\
(-c_{\nu23}s_{\nu12}-s_{\nu23}c_{\nu23}s_{\nu13} e^{i\delta_{\nu}})e^{-i\beta}
&c_{\nu23}c_{\nu12}-s_{\nu23}s_{\nu12}s_{\nu13} e^{i\delta_{\nu}} 
&s_{\nu23}c_{\nu13}e^{i(\gamma-\beta)} \\
(s_{\nu23}s_{\nu12}-c_{\nu23}c_{\nu12}s_{\nu13} e^{i\delta_{\nu}})e^{-i\gamma}
 & (-s_{\nu23}c_{\nu12}-c_{\nu23}s_{\nu12}s_{\nu13} 
e^{i\delta_{\nu}})e^{-i(\gamma-\beta)} 
& c_{\nu23}c_{\nu13}\\ 
\end{array}
\right) \ .\nonumber\\
& &
\label{majorana}
\end{eqnarray}
Here, \(\alpha_i^e\) comes from the rephasing in the charged lepton fields 
to make the choice of phase convention.
The CP-violating phase \(\delta_{\nu}\), the additional Majorana 
phase $\beta$ and $\gamma$ \cite{bilenky,Doi} 
in the representation Eq.~(\ref{majorana}) are calculable and obtained as
\begin{eqnarray}
\delta_\nu  &=& \mbox{arg}
          \left[
             \frac{U_{12}U_{22}^*}{U_{13}U_{23}^*} + 
             \frac{|U_{12}|^2}{1-|U_{13}|^2}
          \right] 
 \simeq \mbox{arg} \left(
          \frac{U_{12} U_{22}^*}{U_{13} U_{23}^*} \right) \nonumber \\ 
 &\simeq& \varphi_e-\varphi_{\nu} - \frac{1}{2}(\phi_e-\phi_{\nu}),
 \simeq \left\{
	\begin{array}{ll}
	-\frac{\eta_e}{2}+\eta_\nu-\frac{\pi}{4}  & \quad \mbox{for } 0<\eta_e<\pi \\
	-\frac{\eta_e}{2}+\eta_\nu-\frac{3\pi}{4} & \quad \mbox{for } \pi<\eta_e<2\pi
  	\end{array}
  \right. ,
 \\
%
\beta & =& \mbox{arg} \left( \frac{U_{12}}{U_{11}}\right) 
  \simeq \frac{3\pi}{2}, \\
%
 \gamma  & =& \mbox{arg} \left( \frac{U_{13}}{U_{11}}e^{i\delta_\nu}\right) 
  \simeq \left\{ 
	\begin{array}{ll}
	\frac{\eta_\nu}{2}+ \frac{\pi}{2} & \quad \mbox{for } 0<\eta_e <\pi \\
	\frac{\eta_\nu}{2}- \frac{\pi}{2} & \quad \mbox{for } \pi <\eta_e < 2\pi
	\end{array}
	\right. \ ,
\end{eqnarray}
by using the relation \(m_e \ll m_\tau \).
Hence, we
also predict the averaged neutrino mass 
$\langle m_\nu \rangle$ which appears in the neutrinoless 
double beta decay\cite{Doi} as follows:
\begin{eqnarray}
\langle m_\nu \rangle & \equiv & \left| m_1 U_{11}^2 +m_2 U_{12}^2
+m_3 U_{13}^2 \right| \nonumber \\
 & \simeq & \left| m_1\frac{m_2}{m_2+m_1} + m_2\frac{m_1}{m_2+m_1} e^{3\pi i}\right|
 \ll m_1 \ . \label{NDBD}
\end{eqnarray}
This value of $\langle m_\nu \rangle$ is too small to be observed in near future experiments.
\par
In Fig.~2, we present more detailed predicted values for 
the lepton mixing matrix elements (\(|U_{12}|\), \(|U_{23}|\), and \(|U_{13}|\)), 
and phases (\(\sin\delta_\nu\), \(\sin\beta\), and \(\sin\gamma\)) 
in the $\eta_e$ - $\eta_\nu$ parameter plane. Here we take a value given in Eq.~(\ref{alpha-e}) for the parameter $\alpha_e$ 
and center values given in Eq.~(\ref{neu-mass}) for neutrino masses $m_i$. 

\section{conclusion}
\par 
We have analyzed the lepton mixing matrix of a recently proposed nonsymmetric mass matrix model. 
The model gives a universal description of 
quark and lepton with the same texture form (\ref{ourMf}) 
based on an extended flavor $2 \leftrightarrow 3$ symmetry including a phase $\phi$.
By using the charged lepton masses as inputs, the present model has six adjustable parameters, 
$\alpha_e$, $\eta_e$, $\eta_{\nu}$, $m_1$, $m_2$, and $m_3$   
to reproduce the observed MNSP lepton mixing matrix parameters and neutrino-mass-squared differences. 
We have shown that only the case where the type B assignment for charged leptons and the type A for neutrinos  are taken
can lead to consistent values with neutrino oscillation experiments. 
In this case,  we find that the observed large lepton mixing between the second and third generation 
is realized by a fine tuning of the parameter $\alpha_e$ as given in Eq.~(\ref{alpha-e}). 
It is also shown that the model predicts very small value for $|U_{13}|$, 
which is in contrast to previously proposed model\cite{Koide}\cite{Matsuda2}. 
The $CP$ violating phases \(\delta_\nu\), \(\beta\), and \(\gamma\) in the lepton mixing matrix are obtained.
The decay rate of the neutrinoless double beta decay is also predicted to be almost suppressed.



\begin{acknowledgments}
This work of K.M. was supported by the JSPS, No. 3700.
\end{acknowledgments}

\begin{figure}[htbp]
\begin{center}
\includegraphics{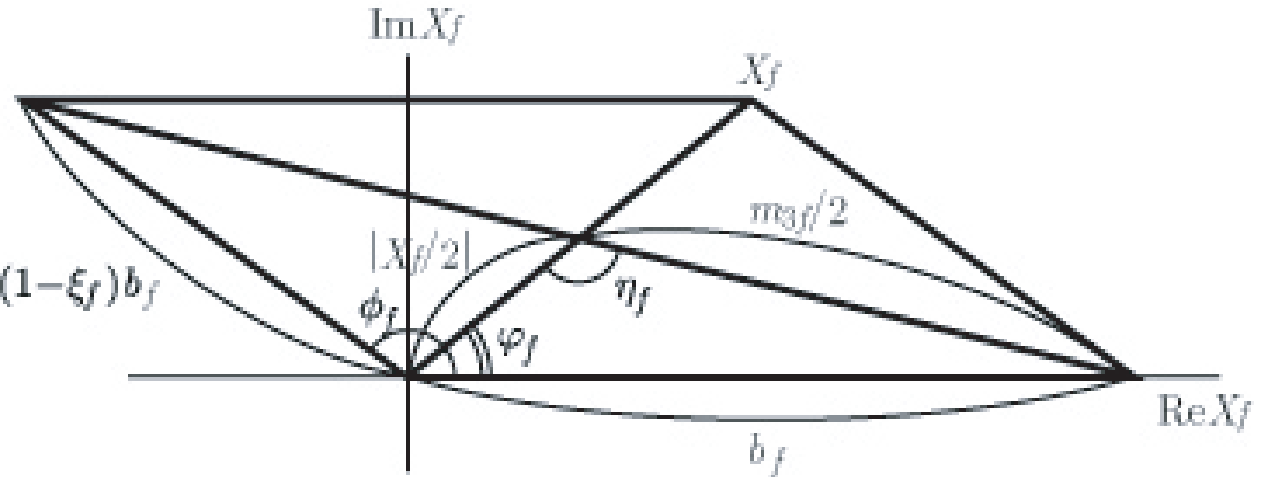}
\end{center}
\caption{%
The relations among the components in the mass matrix, and
the definition of a free $CP$ violating phase parameter $\eta_f$
which is independent of the mass eigenvalues.} 
\label{fig1}
\end{figure}

\begin{figure}[htbp]
\begin{center}
\includegraphics{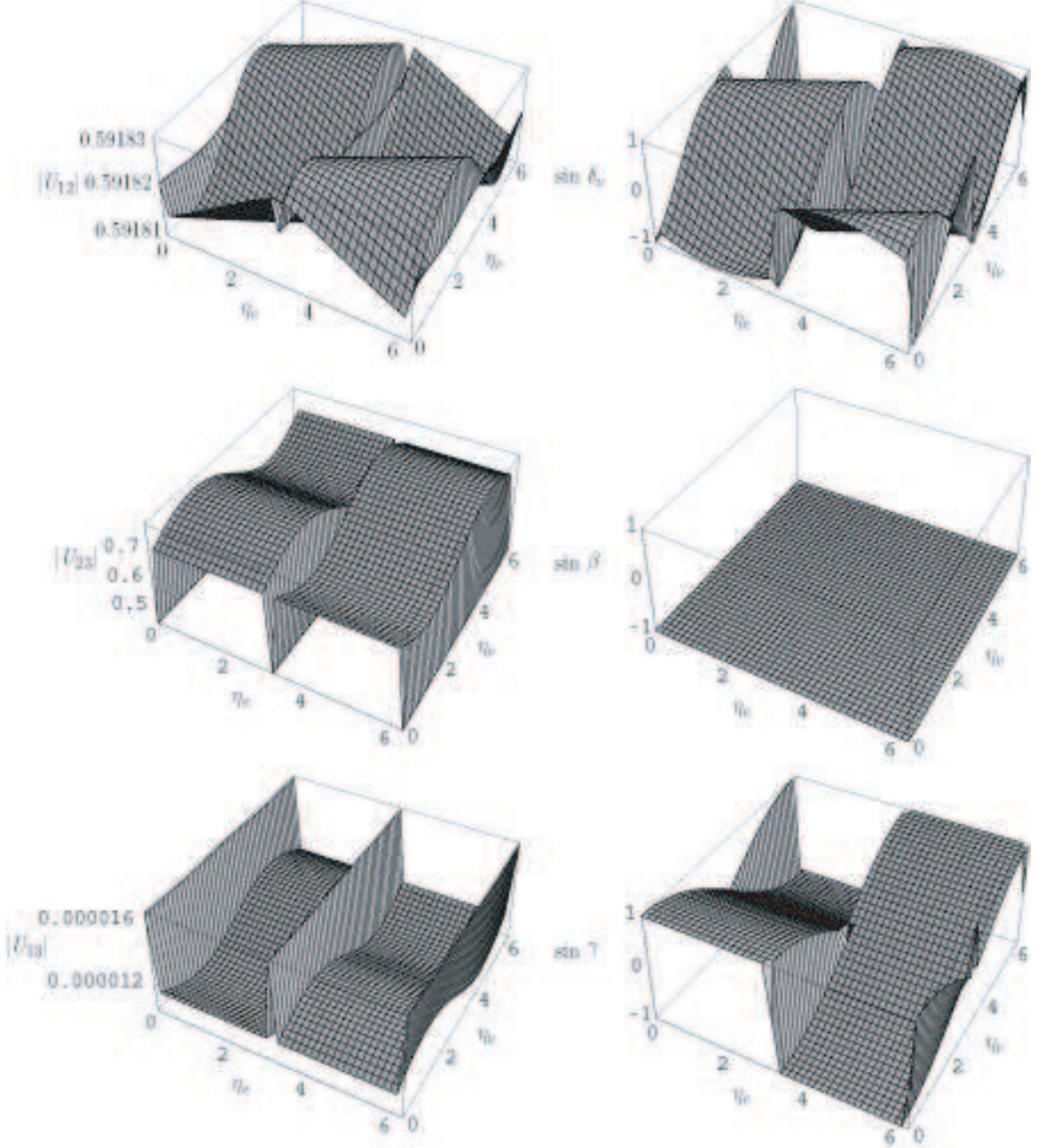}
\end{center}
\caption{%
The predicted values for 
the lepton mixing matrix elements (\(|U_{12}|\), \(|U_{23}|\), and \(|U_{13}|\)), 
and phases (\(\sin\delta_\nu\), \(\sin\beta\), and \(\sin\gamma\)) 
in the $\eta_e$ - $\eta_\nu$ parameter plane 
for the case in which the type B assignment for charged leptons 
and type A for neutrinos are taken.
Here we take a value given in Eq.~(\ref{alpha-e}) for the parameter $\alpha_e$ 
and center values given in Eq.~(\ref{neu-mass}) for neutrino masses $m_i$. }
\label{fig2}
\end{figure}

\end{document}